\documentclass[%
 reprint,
 amsmath,amssymb,
 aps,
prstab,
]{revtex4-2}

\usepackage{graphicx}
\usepackage{dcolumn}
\usepackage{bm}
\usepackage{xcolor} 

\begin{document}

\preprint{APS/123-QED}

\title{Beam shaping using an ultra-high vacuum multileaf collimator and emittance exchange beamline}

\author{N. Majernik$^1$}
\email{NMajernik@g.ucla.edu}

\author{G. Andonian$^1$}
\author{W. Lynn$^1$}

\author{S. Kim$^2$} 

\author{C. Lorch$^1$}

\author{R. Roussel$^3$}

\author{S. Doran$^2$}
\author{E. Wisniewski$^2$}
\author{C. Whiteford$^2$}

\author{P. Piot$^{2,4}$}
\author{J. Power$^2$}
\author{J. B. Rosenzweig$^1$ \\ \phantom{.}}

\affiliation{
$^1$University of California Los Angeles, Los Angeles, California 90095, USA
}

 \affiliation{ 
 $^2$Argonne National Laboratory, Lemont, Illinois 60439, USA
 }
\affiliation{ 
 $^3$SLAC National Accelerator Laboratory, Menlo Park, California 94025, USA
 }

\affiliation{ 
 $^4$Northern Illinois University, DeKalb, Illinois 60115, USA
 }

\date{\today}

\begin{abstract}

We report the development of a multileaf collimator (MLC) for charged particle beams, based on independently actuated tungsten strips which can selectively scatter unwanted particles.
The MLC is used in conjunction with an emittance exchange beamline to rapidly generate highly variable longitudinal bunch profiles.
The developed MLC consists of 40 independent leaves that are 2 mm wide and can move up to 10 mm, and operates in an ultra high vacuum environment, enabled by novel features such as magnetically coupled actuation.
An experiment at the Argonne Wakefield Accelerator, which previously used inflexible, laser-cut masks for beam shaping before an emittance exchange beamline, was conducted to test functionality.
The experiment demonstrated myriad transverse mask silhouettes, as measured on a scintillator downstream of the MLC and the corresponding longitudinal profiles after emittance exchange, as measured using a transverse deflecting cavity.
Rapidly changing between mask shapes enables expeditious execution of various experiments without the downtime associated with traditional methods. 
The many degrees of freedom of the MLC can enable optimization of experimental figures of merit using feed-forward control and advanced machine learning methods.

\end{abstract}

\maketitle


\section{Introduction}

One of the goals in modern accelerator physics is the full control of particle beam distributions in multi-dimensional space \cite{Ha:2022rev}.
Many methods exist for transverse phase space shaping, employing magnetic elements or rigid collimators along the beamline, yet there are fewer reliable options for longitudinal phase space tailoring.
Designer longitudinal profiles of beam current are important in many applications. 
For example, asymmetric (ramped) beam profiles are critical for enhancing efficiency in wakefield-driven acceleration concepts \cite{bane1985collinear,lemery2015tailored} while ramping the beam longitudinal profile in the opposing sense is useful in mitigation of effects stemming from coherent synchrotron radiation \cite{Mitchell:2013}.
Drive beam current profile tailoring is also consequential from the standpoint of enhancing the final energy output in free-electron lasers \cite{Ding:2016}.
 
In recent years, many methods for manipulating the beam longitudinal profile have been experimentally explored. 
Some of these methods introduce, then remove, specific correlations in the beam 6D phase space.
Such beam shaping methods that have been experimentally demonstrated include using higher-order multipole magnets in a dispersive dogleg section \cite{england2008generation}, rigid masking at high dispersion \cite{Muggli:2008}, dual high frequency RF modulations \cite{Piot:2012}, and self-generated wakefield modulations coupled with magnetic compression \cite{andonian2017generation,Antipov:2013}.
In addition, direct laser shaping on the cathode has produced sources with controllable current profiles \cite{loisch2018photocathode}, while inverse free-electron laser interactions have demonstrated bunch train generation at high repetition rates \cite{Sudar:2020}.
Finally, transverse-to-longitudinal emittance exchange (EEX) methods have also successfully produced a variety of beam shapes by design through complex, multi-dimensional phase space manipulations.

\begin{figure*}[t]
   \centering
   \includegraphics[width=\textwidth]{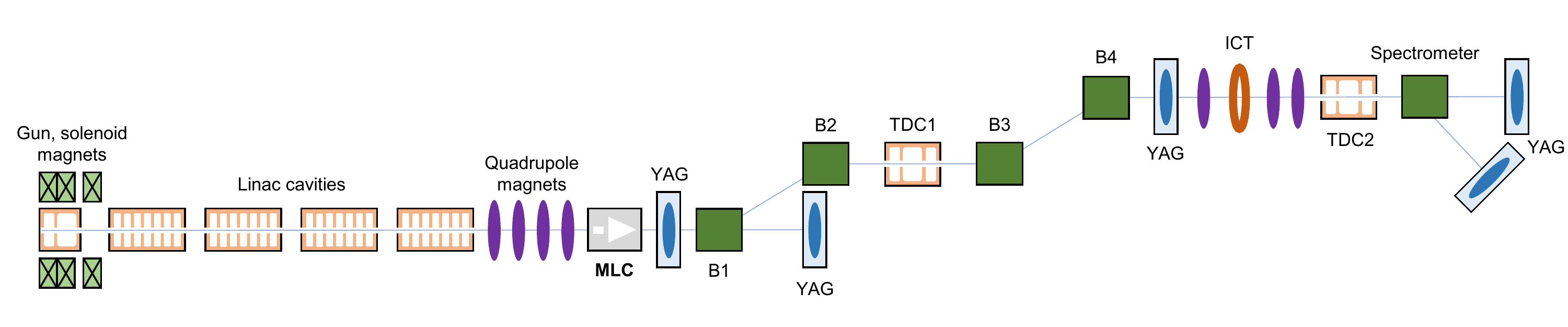}
   \caption{AWA drive linac and EEX beamline (not to scale), adapted from~\cite{ha2017precision} to include the MLC. There are six linac cavities at the AWA, but two cavities were not used in this experimental demonstration.}
   \label{fig:beamlineSchematic}
\end{figure*}

Specifically, transverse distribution masking combined with EEX \cite{Sun:PhysRevLett.105.234801,Ha:2016h} (See Figure \ref{fig:beamlineSchematic}) is a versatile option for shaping the longitudinal profiles of high charge bunches with a high degree of precision. 
In EEX, one of the transverse phase-space planes of the beam is swapped with the longitudinal phase plane. 
EEX is often accomplished by placing a transverse deflecting cavity between two dogleg transport sections~\cite{Emma:PhysRevSTAB.9.100702,RousselThesis}, although other beamline layouts are possible \cite{Xiang:2011}. 
The EEX approach allows for the generation of high-charge bunches with current profiles shaped with a precision that is difficult to achieve using other techniques \cite{Ha:2022rev}.

\begin{figure*}[t]
   \centering
   \includegraphics*[width=0.75\textwidth]{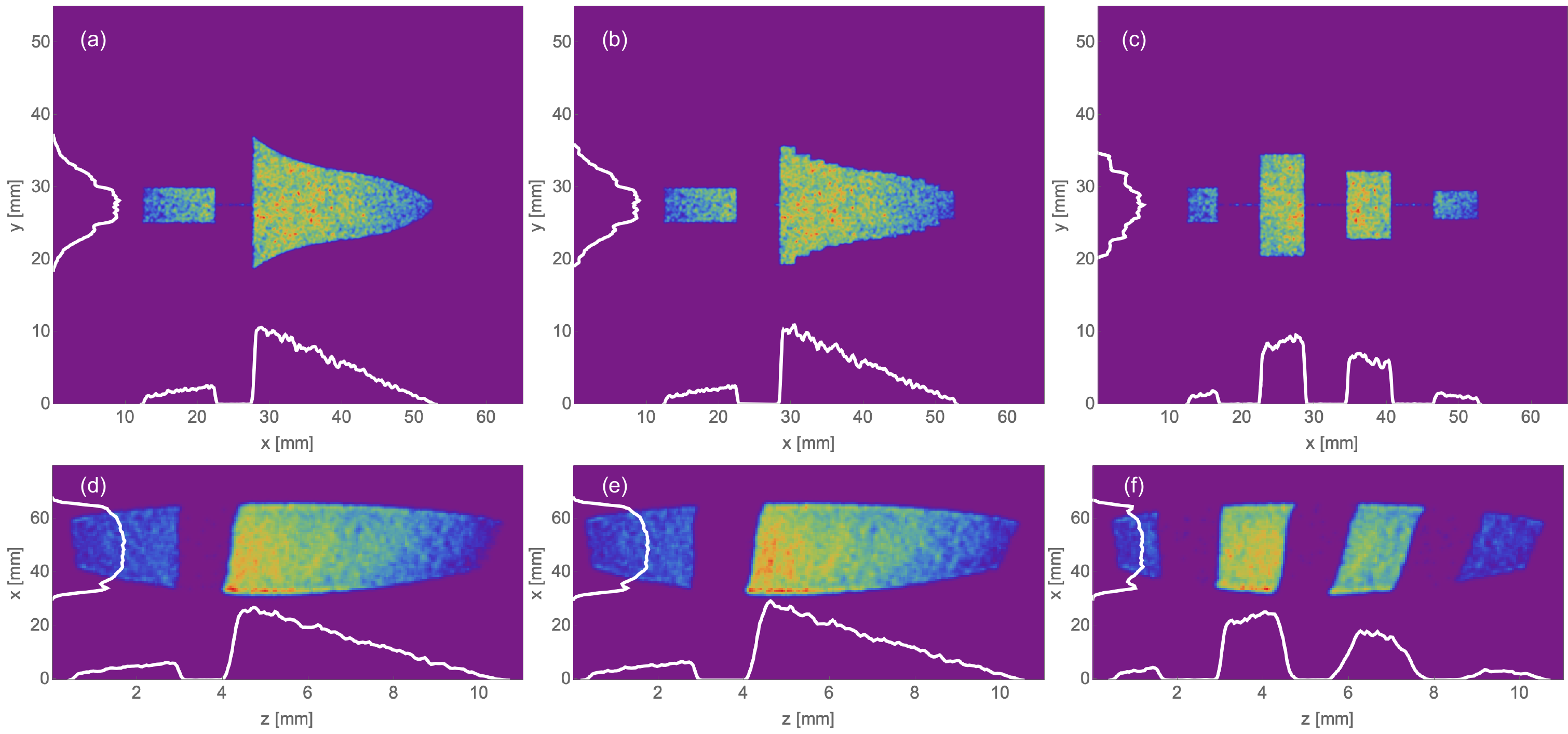}
   \caption{Simulated screen images with vertical and horizontal projections for three mask configurations. The top row \textbf{(a-c)} shows the electron shadow of the mask shortly downstream of the mask while the bottom row \textbf{(d-f)} shows the beam after the EEX beamline and a transverse-deflecting cavity, revealing the longitudinal structure of the beam. The first column shows a smooth mask for a triangle current profile with witness bunch, the second column is the MLC set to approximate the smooth mask, and the final column is a three driver bunch profile with a witness.}
   \label{fig:combinedSimulationFigures}
\end{figure*}

The EEX beamline at the Argonne Wakefield Accelerator Facility (AWA)  has generated electron beams of many different longitudinal profiles \cite{ha2017precision}, and recently used such beams in the demonstration of high transformer ratios in dielectric wakefield acceleration \cite{gao2018observation} and plasma wakefield acceleration \cite{roussel2020PRL}.
The transformer ratio, the ratio of the maximum accelerating field to the maximum decelerating field, $\mathcal{R} \equiv |W_+/W_-|$, is limited to two for longitudinally symmetric bunches \cite{bane1985collinear}. However, using a drive bunch that has an asymmetric current profile -- with a ramp increasing in time followed by a sharp drop in current -- transformer ratios greater than two are achievable and have been demonstrated \cite{loisch2018observation,gao2018observation,roussel2020PRL}.

At the AWA EEX beamline, a linear accelerator (linac) consisting of four 1.3-GHz cavities to accelerate electron bunches up to 43 MeV energy.
A series of quadrupole magnets are then used to control the beam transverse-phase-space parameters at the location of the transverse mask. 
After the beam is propagated through the mask, it traverses the EEX beamline, which consists of four dipoles and a transverse deflecting cavity (TDC) to enable a transformation which swaps the horizontal and longitudinal phase spaces.
The beam charge is monitored in this experiment with an integrating current transformer, and the longitudinal phase space downstream of the EEX beamline is characterized in a diagnostic line, as illustrated in  Figure~\ref{fig:beamlineSchematic}.

The transverse masking has been previously accomplished using laser-cut tungsten masks and changing the resultant longitudinal beam profile required physically changing the mask shape by installing newly cut masks into the UHV beamline.
This is a time-intensive process which has prohibited  quick iteration and dynamical refinement of the beam current profile. 
In this paper, we describe an apparatus that replaces the laser-cut masks with a fully adjustable multi-leaf collimator (MLC) \cite{AACProc, IPACProc}.
The MLC is a device with dozens of independently actuated leaves which intercept the beam trajectory to create a custom aperture \cite{jordan1994design,boyer1992clinical,ge2014toward}. 
MLCs are common in radiotherapy applications to shape the radiation beam to precisely match the shape of the target from any angle, thereby delivering an effective dose while reducing exposure to non-targeted regions. 
In the context of an EEX beamline, the MLC enables rapid, near-arbitrary control over the drive and witness current profiles and spacing.

Start-to-end beam dynamics simulations are performed, to compare beams produced by a practical MLC and existing laser-cut masks, illustrating operational equivalence between the approaches. 
The engineering design, fabrication, and assembly of a forty leaf, ultra-high vacuum compatible MLC are presented. This implementation includes a number of novel approaches which enable unique new capabilities.
Finally, an experiment was conducted to shape and characterize various profile beams using the MLC and EEX beamline at the AWA. The results of the MLC-based beam shaping are analyzed.  The measurements show both the rapid manipulation of the beam profile and control over fine features of the longitudinal distribution.

\section{Simulations}

Unlike a monolithic, rigid, laser-cut mask, the MLC is composed of discrete elements, resulting in a `pixelation' of the masked beam. It is necessary to establish that an MLC can reproduce current profiles of interest with sufficient fidelity to be useful for experiments such as high transformer ratio wakefield acceleration.
In this regard, the desired beam profile is ramped, \textit{i.e.} triangle shaped, or a bunch train, and includes a trailing witness beam to sample the wakefield.
To that end, beam dynamics simulations of the AWA EEX beamline using OPAL \cite{adelmann2009object} have been conducted which compare the performance of laser-cut masks to their approximations by a MLC.

For the tracking simulations, all relevant collective effects such as space-charge and coherent synchrotron radiation were included.
Incident beam parameters are optimized to maintain the shaped distribution despite higher-order aberrations and collective effects~\cite{PhysRevAccelBeams.19.121301}.
The simulated MLC has 40 leaves, 20 per side, with a leaf width of 2 mm and a travel length of 10 mm for a maximum open aperture of 40 mm $\times$ 20 mm. 
The panels in Figure~\ref{fig:combinedSimulationFigures} show the shadow of a smooth, laser-cut mask and MLC, set to approximate the same profile, respectively on a transverse diagnostic  screen shortly downstream of the mask. The discrete nature of the MLC mask is clearly evident as a stair-step feature that approximates the smooth shape. 
These beams were virtually propagated using the simulation model through the EEX beamline (Figure \ref{fig:beamlineSchematic}) and then through the streaking diagnostic which utilizes a transverse deflecting cavity (TDC). This  reveals the current profiles resulting from the mask and EEX implementation.

The EEX beamline is designed for a magnification factor of 4:1 when going from the horizontal to longitudinal coordinate. The final current profiles are shown in Figures \ref{fig:combinedSimulationFigures}. The agreement between the MLC and laser-cut mask current profiles makes it evident that a practical MLC is functionally equivalent to laser-cut masks in regard to the output beam profile, yet the MLC provides greater flexibility for rapid, nearly arbitrary beam shaping.

In addition to a single drive beam with a trailing witness beam, Fig.~\ref{fig:combinedSimulationFigures} shows a beam distribution consisting of three drive bunchlets that is followed by a witness beam, thus creating the possibility of resonant wakefield excitation.
The final distribution after the EEX beamline is also shown.
The effective equivalence between the shapes from a smooth mask and those from the MLC illustrate the benefit of using a configurable mask for rapid turnover without sacrificing high precision.

\section{Design and fabrication}

\begin{figure}[h]
   \centering
   \includegraphics*[width=0.65\columnwidth]{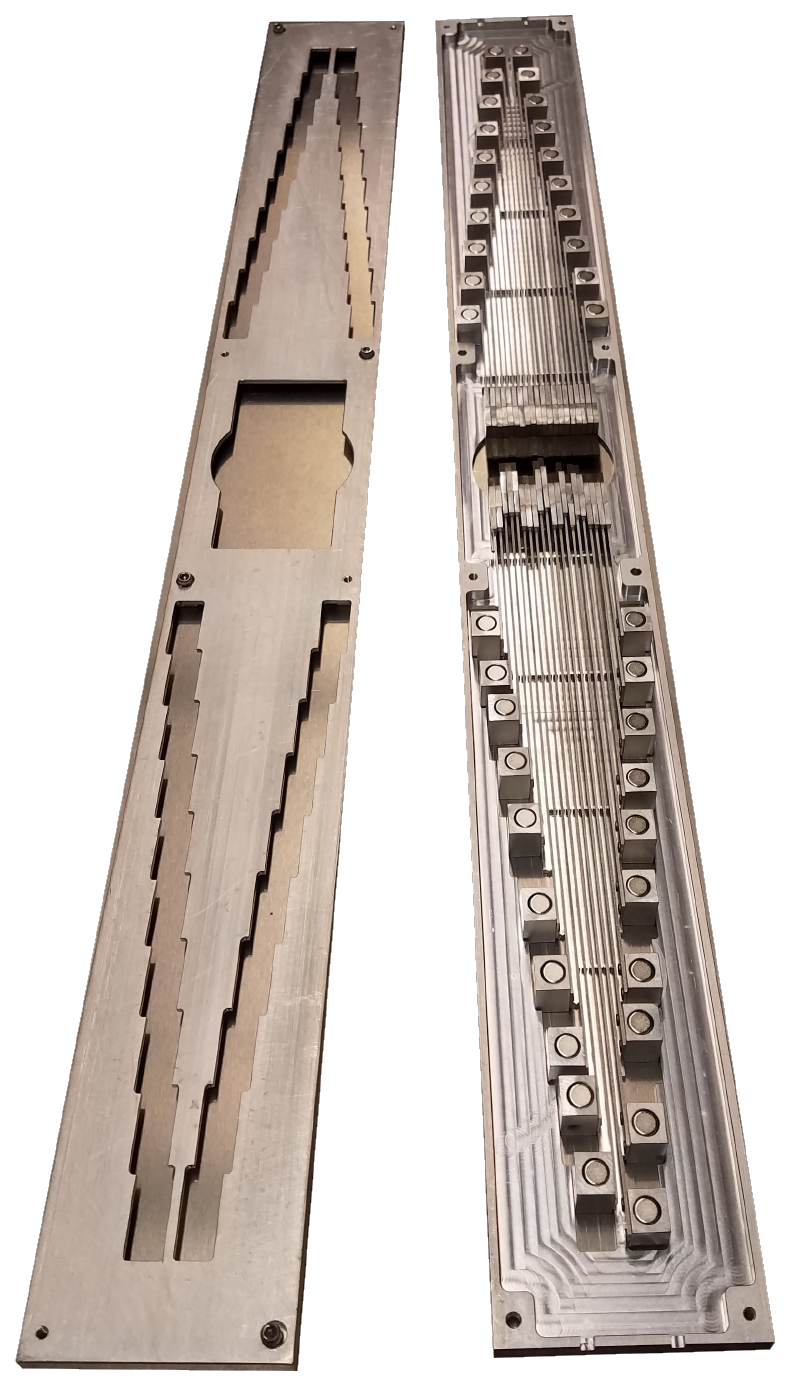}
   \caption{Photograph of the MLC endoskeleton. Forty leaves are installed. The left half is attached on top of the right half to form the cassette for insertion into the vacuum chamber.}
   \label{fig:OpenCassette}
\end{figure}

\begin{figure}[h]
   \centering
   \includegraphics[width=0.85\columnwidth]{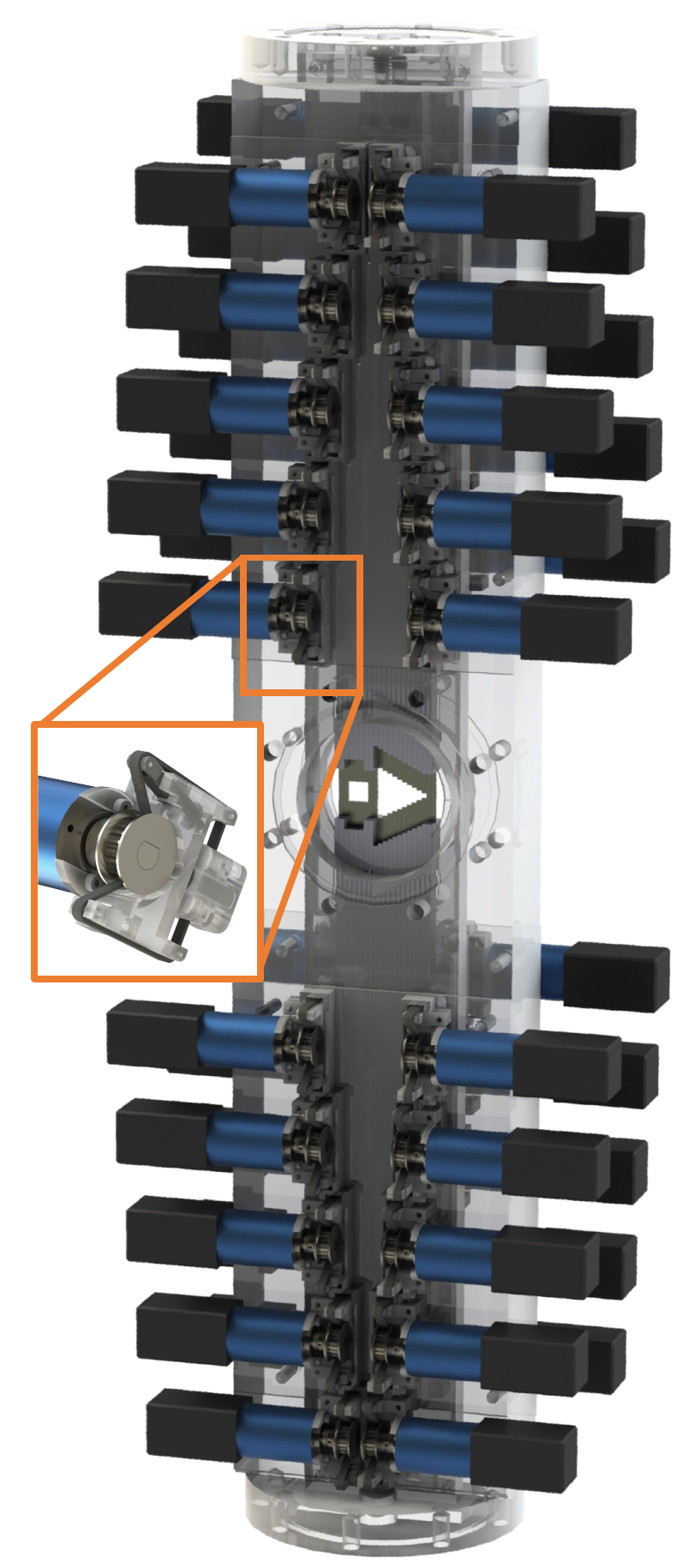}
   \caption{A render of the multileaf collimator with the vacuum chamber and some support structures made translucent for clarity, shown here creating a mask for a ramped drive beam and witness beam. Inset shows a drivetrain module (Compare to Figure \ref{fig:drivetrainPhoto}).}
   \label{fig:completeRender}
\end{figure}

\begin{figure}[h]
   \centering
   \includegraphics*[width=0.75\columnwidth]{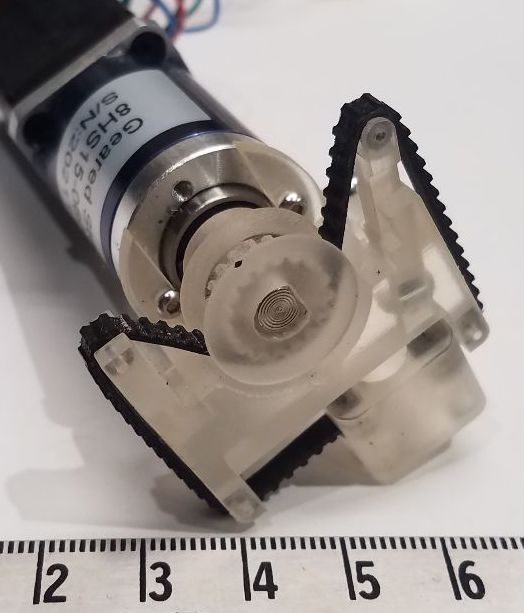}
   \caption{Photograph of an assembled, belt-actuated drivetrain module, shown with a ruler with 1 cm graduations.}
   \label{fig:drivetrainPhoto}
\end{figure}

\begin{figure*}[t] 
   \centering
   \includegraphics*[width=\textwidth]{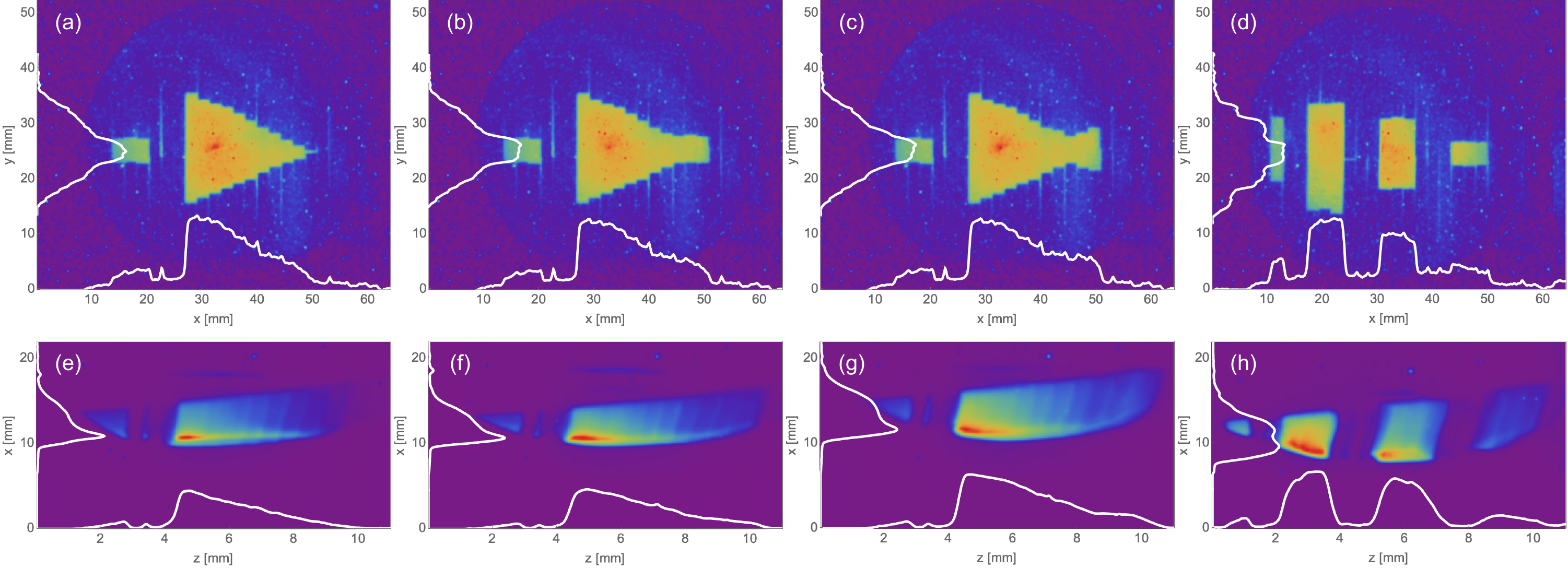}
   \caption{Screen images with vertical and horizontal projections for four MLC mask configurations. {\bf Top row (a-d):} electron shadow of the mask 0.28~m downstream of the MLC for drive and witness beam distributions, for triangle, triangle with doorstep, triangle with bowtie, and three bunch profile respectively. White lines show the transverse projections of the beam. {\bf Bottom row (e-h):}  Beam images after propagating the EEX beamline and a transverse deflecting cavity, for triangle, triangle with doorstep, triangle with bowtie, and three bunch profile respectively.}
   \label{fig:combinedScreenImages}
\end{figure*}

There are many differences between commercially available MLCs for radiotherapy and the experimental requirements for EEX masking, but the most significant is the need for UHV compatibility. 
The main design principle for the UHV-compatible MLC (Figure \ref{fig:completeRender}) relies on magnetically coupling each of the 40 leaves to the exterior of the vacuum vessel. The viability of this concept requires that the magnets on the exterior of the vacuum vessel remain coupled to the magnets on the interior. However, there is a tradeoff between magnet compactness and coupling strength. 

The coupling strength should exceed the leaf weight, magnet-to-chamber friction, other friction sources, and have some safety margin. Bench top tests were conducted to find an acceptable solution \cite{AACProc}. Each of the 40 leaves is composed of a nickel-copper-nickel plated N52 neodymium (NdFeB) permanent magnet, with a 6.4 mm diameter and 12.7 mm length, connected to an aluminum pusher rod. Each pusher rod is, in turn, connected to a 1 mm thick, 1.96 mm wide tungsten tip which interacts with the electron beam. This tip has a tapered profile, cut with multi-axis wire electric discharge machining (EDM), which allows it to be press fit into the aluminum pusher rod. Due to the stringent vacuum requirements and contaminant control at the AWA, no adhesives, lubricants, or polymers were used in the construction of the in vacuum elements of the MLC. The leaf assemblies are held in an aluminum endoskeleton (Figure \ref{fig:OpenCassette}) forming a single cassette for installation. The endoskeleton incorporates small pin features which prevent the leaves from rubbing each other. 

The vacuum vessel was machined from a single piece of 304 stainless steel. The sides of the vessel have pockets to minimize the material thickness between the magnets to enhance the magnetic coupling. Finite element analysis (FEA) simulations were performed to ensure that the thin walls would not excessively deflect from atmospheric pressure while under vacuum. As a compromise to optimize coupling strength while maintaining vacuum integrity,  a thickness of 1.5~mm was selected. The vacuum vessel has a high precision, high aspect ratio (12.7 mm $\times$ 63.5 mm $\times$ 590 mm) wire EDM'd slot to accept the cassette. 

Early designs of the MLC envisioned the use of cable actuation with a spring-driven return to move the external magnets which, in turn, moves the leaves. This approach suffered from issues arising from stick-slip motion. Ultimately, a modular drive-train (Figure \ref{fig:drivetrainPhoto}), was designed. The modular approach is driven bi-directionally by a micro-timing belt, eliminating the reliability and reproducibility issues arising from the spring return. Further, since the design is modular, each of the 40 channels has an identical drive-train, which can be individually assembled and dropped into place, eliminating the cable routing and improving consistency between channels. 
By incorporating a belt tensioning arm and using ball bearings rather than static surfaces, the motion is consistent over time as well. The drivetrain module design and realization is heavily leveraged off of 3D printing which provides precise and economical components. This, combined with the serpentine belt layout, ensures that the new drivetrain module can be tightly tiled while still driving all 40 MLC leaves. This is a crucial consideration since inefficient tilings would increase the required vacuum vessel size, leading to added expense and complications.

The magnetic coupling and external actuation are a UHV safe and economical solution, but the ability to dead-reckon the positions of the leaves is lost since there is no rigid connection to a position encoder. Determining the leaf position therefore requires another form of feedback. For precise read-out of the leaf positions, the YAG screen images, like the simulations of Figure~\ref{fig:combinedSimulationFigures} can be used, but this measurement is destructive to the beam. To supplement this diagnostic, a camera can be used to view the MLC leaves, so the leaf positions can be determined in a fully online, nondestructive fashion.

\section{Experimental results}

After UHV cleaning, the MLC was installed in the AWA EEX beamline. An experiment was conducted with the goal of generating beams with various longitudinal profiles which demonstrate well the range of variation in beam profile shaping. 
Different configurations of the MLC leaves created various masks which transversely shape the beam in diverse ways. 
The transformation into longitudinal profiles is then accomplished via the EEX beamline. The beam parameters for this experimental run include a charge of 5 nC, central energy of 43 MeV. The beam is expanded (FWHM$_x$ = 31 mm, FWHM$_y$ = 18 mm) to  overfill the nominal aperture at the MLC.
For each configuration, the electron beam was passed through the MLC and imaged on a YAG:Ce scintillator screen 28 cm downstream, with no intervening electron optics. 
The masked beam,  after removal of the YAG scintillator, was then propagated through the EEX beamline. A second transverse deflecting cavity, in addition to the cavity in the EEX beamline, is used as a diagnostic of the current profile of the masked beam.

Four examples of masks producing drive and witness bunches are shown in Figure \ref{fig:combinedScreenImages}. These images demonstrate the ability of the MLC to create both a wide range of current profiles, and also to offer precise control over fine features.  Based on the EEX parameters, the MLC's 2 mm wide leaves correspond to features on the $\sim$500 \textmu m scale in the longitudinal coordinate.

\begin{figure}[h]
   \centering
   \includegraphics*[width=\columnwidth]{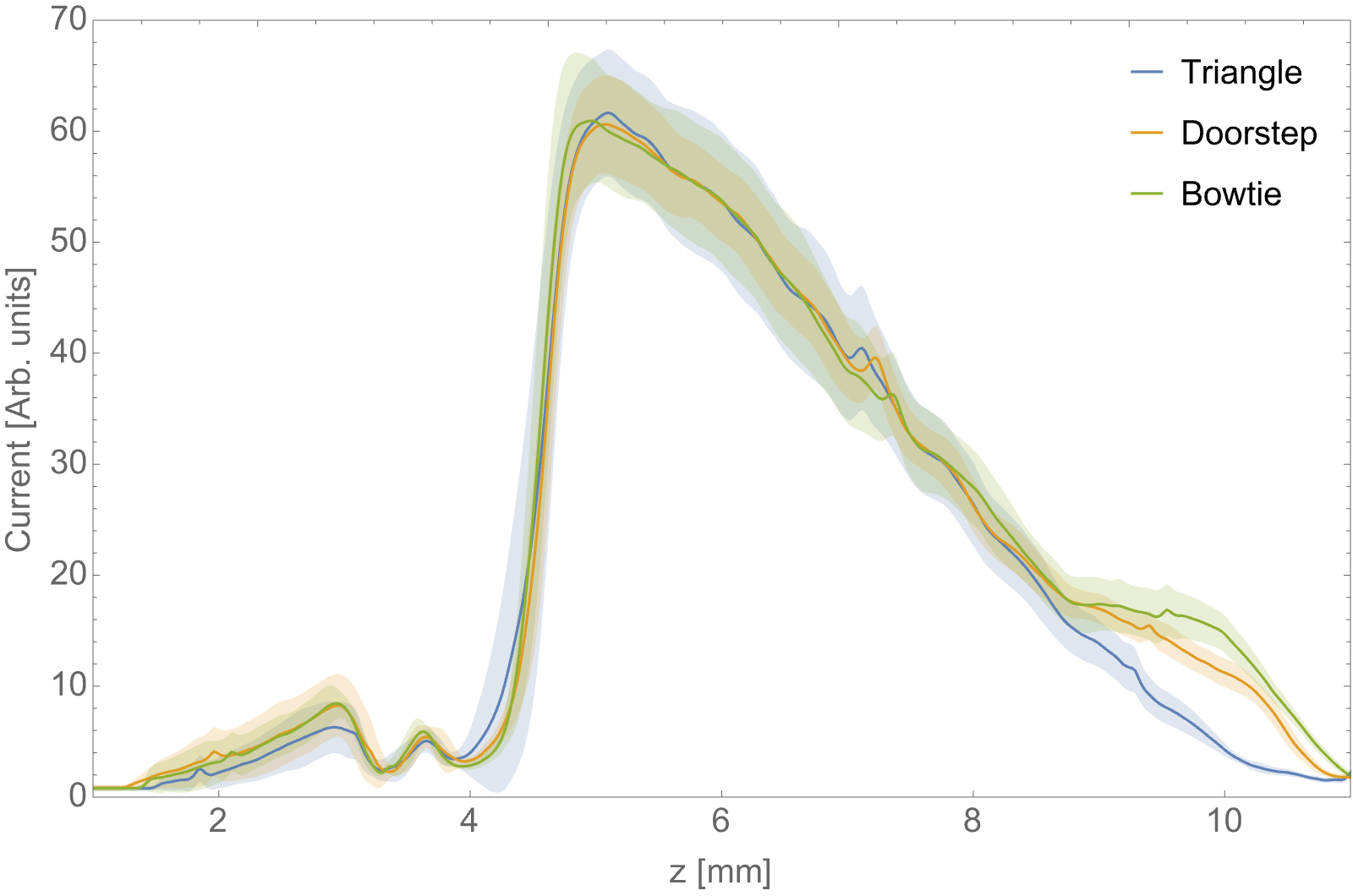}
   \caption{Post-EEX current profiles for the triangle, doorstep, and bowtie masks, shown with shaded bands that correspond to one standard deviation after normalization and alignment. The names in the legend identify the shape of the MLC mask, not necessarily the final beam profile.}
   \label{fig:CurrentProfiles}
\end{figure}

For wakefield drive beams, the head of the beam has a significant impact in optimization for high transformer ratios \cite{bane1985collinear,lemery2015tailored}. 
In Figure~\ref{fig:combinedScreenImages}, we show three drive beams with approximately linear ramped profiles in the beam body and trailing witness beams, but also having three different beam head geometries. The first is a triangle shape, \textit{i.e.} a locally linear ramp, beam with no modification; the second mask generates a beam which includes a corresponding doorstep at the head \cite{bane1985collinear} but, due to the non-uniform transverse pre-collimation beam distribution at the mask, only approximates the doorstep feature after EEX; the third mask, a \emph{bowtie}, increases the gap of the head leaves to compensate for the transverse distribution, ultimately resulting in a doorstep after EEX. 
Finally, to illustrate the flexibility of the MLC, the MLC leaves were configured to represent a bunch train and witness, as may be used for resonant wakefield excitation \cite{Chiadroni2017,BarberThesis,rosenzweig2010plasma} or resonant excitation of coherent Cerenkov radiation \cite{andonian2011resonant}. In this case, a set of three equidistant bunches, followed by a witness beam, was demonstrated.

The current profiles for the triangle, doorstep, and bowtie masks are shown in further detail in Figure \ref{fig:CurrentProfiles}.  The one standard deviation bands show the shot-to-shot jitter of the beamline, attributed to initial condition variations at the beam source and phase error in the RF systems.  
The beams show a steep drop-off at the tail, and significant difference of the features at the head.
The results illustrate that the MLC is capable of precisely controlling the current profiles of beams, at a level relevant for high transformer ratio acceleration.

It is important to discuss the features at the head of the beam. The transverse shadow immediately after the MLC mask does not necessarily correspond to a direct translation into longitudinal projection, due to higher order effects in the EEX transport.
As can be seen for the doorstep mask, the resultant current has a head that is also linearly rising, albeit at a different slope from the bulk of the beam,
In contrast, the bowtie mask leads to a beam head that has a flat profile. However, the ability to dynamically move the leaf of the collimator while measuring on a single-shot basis the current profile will enable the deployment of optimization technique to devise MLC settings that precompensate for distortion arising from nonlinear transport and, possibly, collective effects.

\section{Conclusion}

A UHV-compatible, multi-leaf collimator has been designed, fabricated, and integrated into the AWA EEX beamline, as an alternative to the fixed profile laser-cut tungsten masks previously used. The MLC upgrade permits rapid updates of the driver and witness current profiles, allowing for iterative refinement that is not readily practicable  with a fixed mask system.  
The experimental results from the MLC, including the masked beams and their resultant current profiles, have been shown, including examples corresponding to ramped, high transformer ratio drive beams with trailing witness beams and resonant bunch trains. 
The concept may find use in other accelerator beamlines that rely on transverse masking and require strict UHV levels, for example at the BNL ATF \cite{BarberThesis, Muggli:2008} or at SLAC FACET \cite{yakimenko2019facet}. 
Since the MLC has a large number of system variables for tuning, it is an ideal instrument to pair with machine learning control \cite{scheinker2021extremum, duris2020bayesian}, for the optimization of beam shaping. This approach would enable many applications, including high transformer ratio wakefield acceleration and improved XFEL performance.

\begin{acknowledgments}
This work was supported by the National Science Foundation under Grant No. PHY-1549132 and DOE Grant No. DE-SC0017648. The authors thank E. Holmes, N. Inzunza, S. O'Tool, and N. von Bueren for their assistance assembling components of the MLC. The computing resources used for this research were provided on {\sc bebop}, a high-performance computing cluster operated by the Laboratory Computing Resource Center (LCRC) at ANL.
\end{acknowledgments}

\bibliography{References}

\begin{thebibliography}{35}%
\makeatletter
\providecommand \@ifxundefined [1]{%
 \@ifx{#1\undefined}
}%
\providecommand \@ifnum [1]{%
 \ifnum #1\expandafter \@firstoftwo
 \else \expandafter \@secondoftwo
 \fi
}%
\providecommand \@ifx [1]{%
 \ifx #1\expandafter \@firstoftwo
 \else \expandafter \@secondoftwo
 \fi
}%
\providecommand \natexlab [1]{#1}%
\providecommand \enquote  [1]{``#1''}%
\providecommand \bibnamefont  [1]{#1}%
\providecommand \bibfnamefont [1]{#1}%
\providecommand \citenamefont [1]{#1}%
\providecommand \href@noop [0]{\@secondoftwo}%
\providecommand \href [0]{\begingroup \@sanitize@url \@href}%
\providecommand \@href[1]{\@@startlink{#1}\@@href}%
\providecommand \@@href[1]{\endgroup#1\@@endlink}%
\providecommand \@sanitize@url [0]{\catcode `\\12\catcode `\$12\catcode
  `\&12\catcode `\#12\catcode `\^12\catcode `\_12\catcode `\%12\relax}%
\providecommand \@@startlink[1]{}%
\providecommand \@@endlink[0]{}%
\providecommand \url  [0]{\begingroup\@sanitize@url \@url }%
\providecommand \@url [1]{\endgroup\@href {#1}{\urlprefix }}%
\providecommand \urlprefix  [0]{URL }%
\providecommand \Eprint [0]{\href }%
\providecommand \doibase [0]{https://doi.org/}%
\providecommand \selectlanguage [0]{\@gobble}%
\providecommand \bibinfo  [0]{\@secondoftwo}%
\providecommand \bibfield  [0]{\@secondoftwo}%
\providecommand \translation [1]{[#1]}%
\providecommand \BibitemOpen [0]{}%
\providecommand \bibitemStop [0]{}%
\providecommand \bibitemNoStop [0]{.\EOS\space}%
\providecommand \EOS [0]{\spacefactor3000\relax}%
\providecommand \BibitemShut  [1]{\csname bibitem#1\endcsname}%
\let\auto@bib@innerbib\@empty
\bibitem [{\citenamefont {Ha}\ \emph {et~al.}(2022)\citenamefont {Ha},
  \citenamefont {Kim}, \citenamefont {Power}, \citenamefont {Sun},\ and\
  \citenamefont {Piot}}]{Ha:2022rev}%
  \BibitemOpen
  \bibfield  {author} {\bibinfo {author} {\bibfnamefont {G.}~\bibnamefont
  {Ha}}, \bibinfo {author} {\bibfnamefont {K.~J.}\ \bibnamefont {Kim}},
  \bibinfo {author} {\bibfnamefont {J.~G.}\ \bibnamefont {Power}}, \bibinfo
  {author} {\bibfnamefont {Y.}~\bibnamefont {Sun}},\ and\ \bibinfo {author}
  {\bibfnamefont {P.}~\bibnamefont {Piot}},\ }\bibfield  {title} {\bibinfo
  {title} {{Bunch shaping in electron linear accelerators}},\ }\href@noop {}
  {\bibfield  {journal} {\bibinfo  {journal} {Rev.Mod.Phys.}\ }\textbf
  {\bibinfo {volume} {94}},\ \bibinfo {pages} {025006} (\bibinfo {year}
  {2022})}\BibitemShut {NoStop}%
\bibitem [{\citenamefont {Bane}\ \emph {et~al.}(1985)\citenamefont {Bane},
  \citenamefont {Chen},\ and\ \citenamefont {Wilson}}]{bane1985collinear}%
  \BibitemOpen
  \bibfield  {author} {\bibinfo {author} {\bibfnamefont {K.~L.}\ \bibnamefont
  {Bane}}, \bibinfo {author} {\bibfnamefont {P.}~\bibnamefont {Chen}},\ and\
  \bibinfo {author} {\bibfnamefont {P.~B.}\ \bibnamefont {Wilson}},\
  }\href@noop {} {\emph {\bibinfo {title} {Collinear wake field
  acceleration}}},\ \bibinfo {type} {Tech. Rep.}\ (\bibinfo  {institution}
  {Stanford Linear Accelerator Center},\ \bibinfo {year} {1985})\BibitemShut
  {NoStop}%
\bibitem [{\citenamefont {Lemery}\ and\ \citenamefont
  {Piot}(2015)}]{lemery2015tailored}%
  \BibitemOpen
  \bibfield  {author} {\bibinfo {author} {\bibfnamefont {F.}~\bibnamefont
  {Lemery}}\ and\ \bibinfo {author} {\bibfnamefont {P.}~\bibnamefont {Piot}},\
  }\bibfield  {title} {\bibinfo {title} {Tailored electron bunches with smooth
  current profiles for enhanced transformer ratios in beam-driven
  acceleration},\ }\href@noop {} {\bibfield  {journal} {\bibinfo  {journal}
  {Physical Review Special Topics-Accelerators and Beams}\ }\textbf {\bibinfo
  {volume} {18}},\ \bibinfo {pages} {081301} (\bibinfo {year}
  {2015})}\BibitemShut {NoStop}%
\bibitem [{\citenamefont {Mitchell}\ \emph {et~al.}(2013)\citenamefont
  {Mitchell}, \citenamefont {Qiang},\ and\ \citenamefont
  {Emma}}]{Mitchell:2013}%
  \BibitemOpen
  \bibfield  {author} {\bibinfo {author} {\bibfnamefont {C.~E.}\ \bibnamefont
  {Mitchell}}, \bibinfo {author} {\bibfnamefont {J.}~\bibnamefont {Qiang}},\
  and\ \bibinfo {author} {\bibfnamefont {P.}~\bibnamefont {Emma}},\ }\bibfield
  {title} {\bibinfo {title} {{Longitudinal pulse shaping for the suppression of
  coherent synchrotron radiation-induced emittance growth}},\ }\href@noop {}
  {\bibfield  {journal} {\bibinfo  {journal} {Physical Review Special Topics -
  Accelerators and Beams}\ }\textbf {\bibinfo {volume} {16}},\ \bibinfo {pages}
  {060703} (\bibinfo {year} {2013})}\BibitemShut {NoStop}%
\bibitem [{\citenamefont {Ding}\ \emph {et~al.}(2016)\citenamefont {Ding},
  \citenamefont {Bane}, \citenamefont {Colocho}, \citenamefont {Decker},
  \citenamefont {Emma}, \citenamefont {Frisch}, \citenamefont {Guetg},
  \citenamefont {Huang}, \citenamefont {Iverson}, \citenamefont {Krzywinski},
  \citenamefont {Loos}, \citenamefont {Lutman}, \citenamefont {Maxwell},
  \citenamefont {Nuhn}, \citenamefont {Ratner}, \citenamefont {Turner},
  \citenamefont {Welch},\ and\ \citenamefont {Zhou}}]{Ding:2016}%
  \BibitemOpen
  \bibfield  {author} {\bibinfo {author} {\bibfnamefont {Y.}~\bibnamefont
  {Ding}}, \bibinfo {author} {\bibfnamefont {K.~L.~F.}\ \bibnamefont {Bane}},
  \bibinfo {author} {\bibfnamefont {W.}~\bibnamefont {Colocho}}, \bibinfo
  {author} {\bibfnamefont {F.~J.}\ \bibnamefont {Decker}}, \bibinfo {author}
  {\bibfnamefont {P.}~\bibnamefont {Emma}}, \bibinfo {author} {\bibfnamefont
  {J.}~\bibnamefont {Frisch}}, \bibinfo {author} {\bibfnamefont {M.~W.}\
  \bibnamefont {Guetg}}, \bibinfo {author} {\bibfnamefont {Z.}~\bibnamefont
  {Huang}}, \bibinfo {author} {\bibfnamefont {R.}~\bibnamefont {Iverson}},
  \bibinfo {author} {\bibfnamefont {J.}~\bibnamefont {Krzywinski}}, \bibinfo
  {author} {\bibfnamefont {H.}~\bibnamefont {Loos}}, \bibinfo {author}
  {\bibfnamefont {A.}~\bibnamefont {Lutman}}, \bibinfo {author} {\bibfnamefont
  {T.~J.}\ \bibnamefont {Maxwell}}, \bibinfo {author} {\bibfnamefont {H.-D.}\
  \bibnamefont {Nuhn}}, \bibinfo {author} {\bibfnamefont {D.}~\bibnamefont
  {Ratner}}, \bibinfo {author} {\bibfnamefont {J.}~\bibnamefont {Turner}},
  \bibinfo {author} {\bibfnamefont {J.}~\bibnamefont {Welch}},\ and\ \bibinfo
  {author} {\bibfnamefont {F.}~\bibnamefont {Zhou}},\ }\bibfield  {title}
  {\bibinfo {title} {{Beam shaping to improve the free-electron laser
  performance at the Linac Coherent Light Source}},\ }\href@noop {} {\bibfield
  {journal} {\bibinfo  {journal} {Physical Review Accelerators and Beams}\
  }\textbf {\bibinfo {volume} {19}},\ \bibinfo {pages} {100703} (\bibinfo
  {year} {2016})}\BibitemShut {NoStop}%
\bibitem [{\citenamefont {England}\ \emph {et~al.}(2008)\citenamefont
  {England}, \citenamefont {Rosenzweig},\ and\ \citenamefont
  {Travish}}]{england2008generation}%
  \BibitemOpen
  \bibfield  {author} {\bibinfo {author} {\bibfnamefont {R.}~\bibnamefont
  {England}}, \bibinfo {author} {\bibfnamefont {J.}~\bibnamefont
  {Rosenzweig}},\ and\ \bibinfo {author} {\bibfnamefont {G.}~\bibnamefont
  {Travish}},\ }\bibfield  {title} {\bibinfo {title} {Generation and
  measurement of relativistic electron bunches characterized by a linearly
  ramped current profile},\ }\href@noop {} {\bibfield  {journal} {\bibinfo
  {journal} {Physical review letters}\ }\textbf {\bibinfo {volume} {100}},\
  \bibinfo {pages} {214802} (\bibinfo {year} {2008})}\BibitemShut {NoStop}%
\bibitem [{\citenamefont {Muggli}\ \emph {et~al.}(2008)\citenamefont {Muggli},
  \citenamefont {Yakimenko}, \citenamefont {Babzien}, \citenamefont {Kallos},\
  and\ \citenamefont {Kusche}}]{Muggli:2008}%
  \BibitemOpen
  \bibfield  {author} {\bibinfo {author} {\bibfnamefont {P.}~\bibnamefont
  {Muggli}}, \bibinfo {author} {\bibfnamefont {V.}~\bibnamefont {Yakimenko}},
  \bibinfo {author} {\bibfnamefont {M.}~\bibnamefont {Babzien}}, \bibinfo
  {author} {\bibfnamefont {E.}~\bibnamefont {Kallos}},\ and\ \bibinfo {author}
  {\bibfnamefont {K.~P.}\ \bibnamefont {Kusche}},\ }\bibfield  {title}
  {\bibinfo {title} {{Generation of Trains of Electron Microbunches with
  Adjustable Subpicosecond Spacing}},\ }\href@noop {} {\bibfield  {journal}
  {\bibinfo  {journal} {Physical Review Letters}\ }\textbf {\bibinfo {volume}
  {101}},\ \bibinfo {pages} {4} (\bibinfo {year} {2008})}\BibitemShut {NoStop}%
\bibitem [{\citenamefont {Piot}\ \emph {et~al.}(2012)\citenamefont {Piot},
  \citenamefont {Behrens}, \citenamefont {Gerth}, \citenamefont {Dohlus},
  \citenamefont {Lemery}, \citenamefont {Mihalcea}, \citenamefont {Stoltz},\
  and\ \citenamefont {Vogt}}]{Piot:2012}%
  \BibitemOpen
  \bibfield  {author} {\bibinfo {author} {\bibfnamefont {P.}~\bibnamefont
  {Piot}}, \bibinfo {author} {\bibfnamefont {C.}~\bibnamefont {Behrens}},
  \bibinfo {author} {\bibfnamefont {C.}~\bibnamefont {Gerth}}, \bibinfo
  {author} {\bibfnamefont {M.}~\bibnamefont {Dohlus}}, \bibinfo {author}
  {\bibfnamefont {F.}~\bibnamefont {Lemery}}, \bibinfo {author} {\bibfnamefont
  {D.}~\bibnamefont {Mihalcea}}, \bibinfo {author} {\bibfnamefont
  {P.}~\bibnamefont {Stoltz}},\ and\ \bibinfo {author} {\bibfnamefont
  {M.}~\bibnamefont {Vogt}},\ }\bibfield  {title} {\bibinfo {title}
  {{Generation and Characterization of Electron Bunches with Ramped Current
  Profiles in a Dual-Frequency Superconducting Linear Accelerator}},\
  }\href@noop {} {\bibfield  {journal} {\bibinfo  {journal} {Physical Review
  Letters}\ }\textbf {\bibinfo {volume} {108}},\ \bibinfo {pages} {034801}
  (\bibinfo {year} {2012})}\BibitemShut {NoStop}%
\bibitem [{\citenamefont {Andonian}\ \emph {et~al.}(2017)\citenamefont
  {Andonian}, \citenamefont {Barber}, \citenamefont {O’Shea}, \citenamefont
  {Fedurin}, \citenamefont {Kusche}, \citenamefont {Swinson},\ and\
  \citenamefont {Rosenzweig}}]{andonian2017generation}%
  \BibitemOpen
  \bibfield  {author} {\bibinfo {author} {\bibfnamefont {G.}~\bibnamefont
  {Andonian}}, \bibinfo {author} {\bibfnamefont {S.}~\bibnamefont {Barber}},
  \bibinfo {author} {\bibfnamefont {F.}~\bibnamefont {O’Shea}}, \bibinfo
  {author} {\bibfnamefont {M.}~\bibnamefont {Fedurin}}, \bibinfo {author}
  {\bibfnamefont {K.}~\bibnamefont {Kusche}}, \bibinfo {author} {\bibfnamefont
  {C.}~\bibnamefont {Swinson}},\ and\ \bibinfo {author} {\bibfnamefont
  {J.}~\bibnamefont {Rosenzweig}},\ }\bibfield  {title} {\bibinfo {title}
  {Generation of ramped current profiles in relativistic electron beams using
  wakefields in dielectric structures},\ }\href@noop {} {\bibfield  {journal}
  {\bibinfo  {journal} {Physical review letters}\ }\textbf {\bibinfo {volume}
  {118}},\ \bibinfo {pages} {054802} (\bibinfo {year} {2017})}\BibitemShut
  {NoStop}%
\bibitem [{\citenamefont {Antipov}\ \emph {et~al.}(2013)\citenamefont
  {Antipov}, \citenamefont {Babzien}, \citenamefont {Jing}, \citenamefont
  {Fedurin}, \citenamefont {Gai}, \citenamefont {Kanareykin}, \citenamefont
  {Kusche}, \citenamefont {Yakimenko},\ and\ \citenamefont
  {Zholents}}]{Antipov:2013}%
  \BibitemOpen
  \bibfield  {author} {\bibinfo {author} {\bibfnamefont {S.}~\bibnamefont
  {Antipov}}, \bibinfo {author} {\bibfnamefont {M.}~\bibnamefont {Babzien}},
  \bibinfo {author} {\bibfnamefont {C.}~\bibnamefont {Jing}}, \bibinfo {author}
  {\bibfnamefont {M.}~\bibnamefont {Fedurin}}, \bibinfo {author} {\bibfnamefont
  {W.}~\bibnamefont {Gai}}, \bibinfo {author} {\bibfnamefont {A.}~\bibnamefont
  {Kanareykin}}, \bibinfo {author} {\bibfnamefont {K.}~\bibnamefont {Kusche}},
  \bibinfo {author} {\bibfnamefont {V.}~\bibnamefont {Yakimenko}},\ and\
  \bibinfo {author} {\bibfnamefont {A.}~\bibnamefont {Zholents}},\ }\bibfield
  {title} {\bibinfo {title} {{Subpicosecond Bunch Train Production for a
  Tunable mJ Level THz Source}},\ }\href@noop {} {\bibfield  {journal}
  {\bibinfo  {journal} {Physical Review Letters}\ }\textbf {\bibinfo {volume}
  {111}},\ \bibinfo {pages} {134802} (\bibinfo {year} {2013})}\BibitemShut
  {NoStop}%
\bibitem [{\citenamefont {Loisch}\ \emph
  {et~al.}(2018{\natexlab{a}})\citenamefont {Loisch}, \citenamefont {Good},
  \citenamefont {Gross}, \citenamefont {Huck}, \citenamefont {Isaev},
  \citenamefont {Krasilnikov}, \citenamefont {Lishilin}, \citenamefont
  {Oppelt}, \citenamefont {Renier}, \citenamefont {Stephan} \emph
  {et~al.}}]{loisch2018photocathode}%
  \BibitemOpen
  \bibfield  {author} {\bibinfo {author} {\bibfnamefont {G.}~\bibnamefont
  {Loisch}}, \bibinfo {author} {\bibfnamefont {J.}~\bibnamefont {Good}},
  \bibinfo {author} {\bibfnamefont {M.}~\bibnamefont {Gross}}, \bibinfo
  {author} {\bibfnamefont {H.}~\bibnamefont {Huck}}, \bibinfo {author}
  {\bibfnamefont {I.}~\bibnamefont {Isaev}}, \bibinfo {author} {\bibfnamefont
  {M.}~\bibnamefont {Krasilnikov}}, \bibinfo {author} {\bibfnamefont
  {O.}~\bibnamefont {Lishilin}}, \bibinfo {author} {\bibfnamefont
  {A.}~\bibnamefont {Oppelt}}, \bibinfo {author} {\bibfnamefont
  {Y.}~\bibnamefont {Renier}}, \bibinfo {author} {\bibfnamefont
  {F.}~\bibnamefont {Stephan}}, \emph {et~al.},\ }\bibfield  {title} {\bibinfo
  {title} {Photocathode laser based bunch shaping for high transformer ratio
  plasma wakefield acceleration},\ }\href@noop {} {\bibfield  {journal}
  {\bibinfo  {journal} {Nuclear Instruments and Methods in Physics Research
  Section A: Accelerators, Spectrometers, Detectors and Associated Equipment}\
  }\textbf {\bibinfo {volume} {909}},\ \bibinfo {pages} {107} (\bibinfo {year}
  {2018}{\natexlab{a}})}\BibitemShut {NoStop}%
\bibitem [{\citenamefont {Sudar}\ \emph {et~al.}(2020)\citenamefont {Sudar},
  \citenamefont {Musumeci}, \citenamefont {Ovodenko}, \citenamefont {Murokh},
  \citenamefont {Polyanskiy}, \citenamefont {Pogorelsky}, \citenamefont
  {Fedurin}, \citenamefont {Swinson}, \citenamefont {Kusche}, \citenamefont
  {Babzien},\ and\ \citenamefont {Palmer}}]{Sudar:2020}%
  \BibitemOpen
  \bibfield  {author} {\bibinfo {author} {\bibfnamefont {N.}~\bibnamefont
  {Sudar}}, \bibinfo {author} {\bibfnamefont {P.}~\bibnamefont {Musumeci}},
  \bibinfo {author} {\bibfnamefont {A.}~\bibnamefont {Ovodenko}}, \bibinfo
  {author} {\bibfnamefont {A.}~\bibnamefont {Murokh}}, \bibinfo {author}
  {\bibfnamefont {M.}~\bibnamefont {Polyanskiy}}, \bibinfo {author}
  {\bibfnamefont {I.}~\bibnamefont {Pogorelsky}}, \bibinfo {author}
  {\bibfnamefont {M.}~\bibnamefont {Fedurin}}, \bibinfo {author} {\bibfnamefont
  {C.}~\bibnamefont {Swinson}}, \bibinfo {author} {\bibfnamefont
  {K.}~\bibnamefont {Kusche}}, \bibinfo {author} {\bibfnamefont
  {M.}~\bibnamefont {Babzien}},\ and\ \bibinfo {author} {\bibfnamefont
  {M.}~\bibnamefont {Palmer}},\ }\bibfield  {title} {\bibinfo {title} {{Burst
  mode MHz repetition rate inverse free electron laser acceleration }},\
  }\href@noop {} {\bibfield  {journal} {\bibinfo  {journal} {Physical Review
  Accelerators and Beams}\ }\textbf {\bibinfo {volume} {23}},\ \bibinfo {pages}
  {051301} (\bibinfo {year} {2020})}\BibitemShut {NoStop}%
\bibitem [{\citenamefont {Sun}\ \emph {et~al.}(2010)\citenamefont {Sun},
  \citenamefont {Piot}, \citenamefont {Johnson}, \citenamefont {Lumpkin},
  \citenamefont {Maxwell}, \citenamefont {Ruan},\ and\ \citenamefont
  {Thurman-Keup}}]{Sun:PhysRevLett.105.234801}%
  \BibitemOpen
  \bibfield  {author} {\bibinfo {author} {\bibfnamefont {Y.-E.}\ \bibnamefont
  {Sun}}, \bibinfo {author} {\bibfnamefont {P.}~\bibnamefont {Piot}}, \bibinfo
  {author} {\bibfnamefont {A.}~\bibnamefont {Johnson}}, \bibinfo {author}
  {\bibfnamefont {A.~H.}\ \bibnamefont {Lumpkin}}, \bibinfo {author}
  {\bibfnamefont {T.~J.}\ \bibnamefont {Maxwell}}, \bibinfo {author}
  {\bibfnamefont {J.}~\bibnamefont {Ruan}},\ and\ \bibinfo {author}
  {\bibfnamefont {R.}~\bibnamefont {Thurman-Keup}},\ }\bibfield  {title}
  {\bibinfo {title} {Tunable subpicosecond electron-bunch-train generation
  using a transverse-to-longitudinal phase-space exchange technique},\ }\href
  {https://doi.org/10.1103/PhysRevLett.105.234801} {\bibfield  {journal}
  {\bibinfo  {journal} {Phys. Rev. Lett.}\ }\textbf {\bibinfo {volume} {105}},\
  \bibinfo {pages} {234801} (\bibinfo {year} {2010})}\BibitemShut {NoStop}%
\bibitem [{\citenamefont {Ha}\ \emph {et~al.}(2016{\natexlab{a}})\citenamefont
  {Ha}, \citenamefont {Cho}, \citenamefont {Gai}, \citenamefont {Kim},
  \citenamefont {Namkung},\ and\ \citenamefont {Power}}]{Ha:2016h}%
  \BibitemOpen
  \bibfield  {author} {\bibinfo {author} {\bibfnamefont {G.}~\bibnamefont
  {Ha}}, \bibinfo {author} {\bibfnamefont {M.~H.}\ \bibnamefont {Cho}},
  \bibinfo {author} {\bibfnamefont {W.}~\bibnamefont {Gai}}, \bibinfo {author}
  {\bibfnamefont {K.~J.}\ \bibnamefont {Kim}}, \bibinfo {author} {\bibfnamefont
  {W.}~\bibnamefont {Namkung}},\ and\ \bibinfo {author} {\bibfnamefont {J.~G.}\
  \bibnamefont {Power}},\ }\bibfield  {title} {\bibinfo {title}
  {{Perturbation-minimized triangular bunch for high-transformer ratio using a
  double dogleg emittance exchange beam line}},\ }\href@noop {} {\bibfield
  {journal} {\bibinfo  {journal} {Physical Review Accelerators and Beams}\
  }\textbf {\bibinfo {volume} {19}},\ \bibinfo {pages} {121301} (\bibinfo
  {year} {2016}{\natexlab{a}})}\BibitemShut {NoStop}%
\bibitem [{\citenamefont {Emma}\ \emph {et~al.}(2006)\citenamefont {Emma},
  \citenamefont {Huang}, \citenamefont {Kim},\ and\ \citenamefont
  {Piot}}]{Emma:PhysRevSTAB.9.100702}%
  \BibitemOpen
  \bibfield  {author} {\bibinfo {author} {\bibfnamefont {P.}~\bibnamefont
  {Emma}}, \bibinfo {author} {\bibfnamefont {Z.}~\bibnamefont {Huang}},
  \bibinfo {author} {\bibfnamefont {K.-J.}\ \bibnamefont {Kim}},\ and\ \bibinfo
  {author} {\bibfnamefont {P.}~\bibnamefont {Piot}},\ }\bibfield  {title}
  {\bibinfo {title} {Transverse-to-longitudinal emittance exchange to improve
  performance of high-gain free-electron lasers},\ }\href
  {https://doi.org/10.1103/PhysRevSTAB.9.100702} {\bibfield  {journal}
  {\bibinfo  {journal} {Phys. Rev. ST Accel. Beams}\ }\textbf {\bibinfo
  {volume} {9}},\ \bibinfo {pages} {100702} (\bibinfo {year}
  {2006})}\BibitemShut {NoStop}%
\bibitem [{\citenamefont {Roussel}(2019)}]{RousselThesis}%
  \BibitemOpen
  \bibfield  {author} {\bibinfo {author} {\bibfnamefont {R.}~\bibnamefont
  {Roussel}},\ }\href@noop {} {\emph {\bibinfo {title} {Single-Shot
  Characterization of High Transformer Ratio Wakefields in Nonlinear Plasma}}}\
  (\bibinfo {year} {2019})\BibitemShut {NoStop}%
\bibitem [{\citenamefont {Xiang}\ and\ \citenamefont
  {Chao}(2011)}]{Xiang:2011}%
  \BibitemOpen
  \bibfield  {author} {\bibinfo {author} {\bibfnamefont {D.}~\bibnamefont
  {Xiang}}\ and\ \bibinfo {author} {\bibfnamefont {A.}~\bibnamefont {Chao}},\
  }\bibfield  {title} {\bibinfo {title} {{Emittance and phase space exchange
  for advanced beam manipulation and diagnostics}},\ }\href@noop {} {\bibfield
  {journal} {\bibinfo  {journal} {Physical Review Special Topics - Accelerators
  and Beams}\ }\textbf {\bibinfo {volume} {14}} (\bibinfo {year}
  {2011})}\BibitemShut {NoStop}%
\bibitem [{\citenamefont {Ha}\ \emph {et~al.}(2017)\citenamefont {Ha},
  \citenamefont {Cho}, \citenamefont {Namkung}, \citenamefont {Power},
  \citenamefont {Doran}, \citenamefont {Wisniewski}, \citenamefont {Conde},
  \citenamefont {Gai}, \citenamefont {Liu}, \citenamefont {Whiteford} \emph
  {et~al.}}]{ha2017precision}%
  \BibitemOpen
  \bibfield  {author} {\bibinfo {author} {\bibfnamefont {G.}~\bibnamefont
  {Ha}}, \bibinfo {author} {\bibfnamefont {M.-H.}\ \bibnamefont {Cho}},
  \bibinfo {author} {\bibfnamefont {W.}~\bibnamefont {Namkung}}, \bibinfo
  {author} {\bibfnamefont {J.}~\bibnamefont {Power}}, \bibinfo {author}
  {\bibfnamefont {D.~S.}\ \bibnamefont {Doran}}, \bibinfo {author}
  {\bibfnamefont {E.}~\bibnamefont {Wisniewski}}, \bibinfo {author}
  {\bibfnamefont {M.}~\bibnamefont {Conde}}, \bibinfo {author} {\bibfnamefont
  {W.}~\bibnamefont {Gai}}, \bibinfo {author} {\bibfnamefont {W.}~\bibnamefont
  {Liu}}, \bibinfo {author} {\bibfnamefont {C.}~\bibnamefont {Whiteford}},
  \emph {et~al.},\ }\bibfield  {title} {\bibinfo {title} {Precision control of
  the electron longitudinal bunch shape using an emittance-exchange beam
  line},\ }\href@noop {} {\bibfield  {journal} {\bibinfo  {journal} {Physical
  review letters}\ }\textbf {\bibinfo {volume} {118}},\ \bibinfo {pages}
  {104801} (\bibinfo {year} {2017})}\BibitemShut {NoStop}%
\bibitem [{\citenamefont {Gao}\ \emph {et~al.}(2018)\citenamefont {Gao},
  \citenamefont {Ha}, \citenamefont {Jing}, \citenamefont {Antipov},
  \citenamefont {Power}, \citenamefont {Conde}, \citenamefont {Gai},
  \citenamefont {Chen}, \citenamefont {Shi}, \citenamefont {Wisniewski} \emph
  {et~al.}}]{gao2018observation}%
  \BibitemOpen
  \bibfield  {author} {\bibinfo {author} {\bibfnamefont {Q.}~\bibnamefont
  {Gao}}, \bibinfo {author} {\bibfnamefont {G.}~\bibnamefont {Ha}}, \bibinfo
  {author} {\bibfnamefont {C.}~\bibnamefont {Jing}}, \bibinfo {author}
  {\bibfnamefont {S.}~\bibnamefont {Antipov}}, \bibinfo {author} {\bibfnamefont
  {J.}~\bibnamefont {Power}}, \bibinfo {author} {\bibfnamefont
  {M.}~\bibnamefont {Conde}}, \bibinfo {author} {\bibfnamefont
  {W.}~\bibnamefont {Gai}}, \bibinfo {author} {\bibfnamefont {H.}~\bibnamefont
  {Chen}}, \bibinfo {author} {\bibfnamefont {J.}~\bibnamefont {Shi}}, \bibinfo
  {author} {\bibfnamefont {E.}~\bibnamefont {Wisniewski}}, \emph {et~al.},\
  }\bibfield  {title} {\bibinfo {title} {Observation of high transformer ratio
  of shaped bunch generated by an emittance-exchange beam line},\ }\href@noop
  {} {\bibfield  {journal} {\bibinfo  {journal} {Physical review letters}\
  }\textbf {\bibinfo {volume} {120}},\ \bibinfo {pages} {114801} (\bibinfo
  {year} {2018})}\BibitemShut {NoStop}%
\bibitem [{\citenamefont {Roussel}\ \emph {et~al.}(2020)\citenamefont
  {Roussel}, \citenamefont {Andonian}, \citenamefont {Lynn}, \citenamefont
  {Sanwalka}, \citenamefont {Robles}, \citenamefont {Hansel}, \citenamefont
  {Deng}, \citenamefont {Lawler}, \citenamefont {Rosenzweig}, \citenamefont
  {Ha} \emph {et~al.}}]{roussel2020PRL}%
  \BibitemOpen
  \bibfield  {author} {\bibinfo {author} {\bibfnamefont {R.}~\bibnamefont
  {Roussel}}, \bibinfo {author} {\bibfnamefont {G.}~\bibnamefont {Andonian}},
  \bibinfo {author} {\bibfnamefont {W.}~\bibnamefont {Lynn}}, \bibinfo {author}
  {\bibfnamefont {K.}~\bibnamefont {Sanwalka}}, \bibinfo {author}
  {\bibfnamefont {R.}~\bibnamefont {Robles}}, \bibinfo {author} {\bibfnamefont
  {C.}~\bibnamefont {Hansel}}, \bibinfo {author} {\bibfnamefont
  {A.}~\bibnamefont {Deng}}, \bibinfo {author} {\bibfnamefont {G.}~\bibnamefont
  {Lawler}}, \bibinfo {author} {\bibfnamefont {J.}~\bibnamefont {Rosenzweig}},
  \bibinfo {author} {\bibfnamefont {G.}~\bibnamefont {Ha}}, \emph {et~al.},\
  }\bibfield  {title} {\bibinfo {title} {Single shot characterization of high
  transformer ratio wakefields in nonlinear plasma acceleration},\ }\href@noop
  {} {\bibfield  {journal} {\bibinfo  {journal} {Physical Review Letters}\
  }\textbf {\bibinfo {volume} {124}},\ \bibinfo {pages} {044802} (\bibinfo
  {year} {2020})}\BibitemShut {NoStop}%
\bibitem [{\citenamefont {Loisch}\ \emph
  {et~al.}(2018{\natexlab{b}})\citenamefont {Loisch}, \citenamefont {Asova},
  \citenamefont {Boonpornprasert}, \citenamefont {Brinkmann}, \citenamefont
  {Chen}, \citenamefont {Engel}, \citenamefont {Good}, \citenamefont {Gross},
  \citenamefont {Gr{\"u}ner}, \citenamefont {Huck} \emph
  {et~al.}}]{loisch2018observation}%
  \BibitemOpen
  \bibfield  {author} {\bibinfo {author} {\bibfnamefont {G.}~\bibnamefont
  {Loisch}}, \bibinfo {author} {\bibfnamefont {G.}~\bibnamefont {Asova}},
  \bibinfo {author} {\bibfnamefont {P.}~\bibnamefont {Boonpornprasert}},
  \bibinfo {author} {\bibfnamefont {R.}~\bibnamefont {Brinkmann}}, \bibinfo
  {author} {\bibfnamefont {Y.}~\bibnamefont {Chen}}, \bibinfo {author}
  {\bibfnamefont {J.}~\bibnamefont {Engel}}, \bibinfo {author} {\bibfnamefont
  {J.}~\bibnamefont {Good}}, \bibinfo {author} {\bibfnamefont {M.}~\bibnamefont
  {Gross}}, \bibinfo {author} {\bibfnamefont {F.}~\bibnamefont {Gr{\"u}ner}},
  \bibinfo {author} {\bibfnamefont {H.}~\bibnamefont {Huck}}, \emph {et~al.},\
  }\bibfield  {title} {\bibinfo {title} {Observation of high transformer ratio
  plasma wakefield acceleration},\ }\href@noop {} {\bibfield  {journal}
  {\bibinfo  {journal} {Physical review letters}\ }\textbf {\bibinfo {volume}
  {121}},\ \bibinfo {pages} {064801} (\bibinfo {year}
  {2018}{\natexlab{b}})}\BibitemShut {NoStop}%
\bibitem [{\citenamefont {Majernik}\ \emph
  {et~al.}(2021{\natexlab{a}})\citenamefont {Majernik}, \citenamefont
  {Andonian}, \citenamefont {Roussel}, \citenamefont {Doran}, \citenamefont
  {Ha}, \citenamefont {Power}, \citenamefont {Wisniewski},\ and\ \citenamefont
  {Rosenzweig}}]{AACProc}%
  \BibitemOpen
  \bibfield  {author} {\bibinfo {author} {\bibfnamefont {N.}~\bibnamefont
  {Majernik}}, \bibinfo {author} {\bibfnamefont {G.}~\bibnamefont {Andonian}},
  \bibinfo {author} {\bibfnamefont {R.}~\bibnamefont {Roussel}}, \bibinfo
  {author} {\bibfnamefont {S.}~\bibnamefont {Doran}}, \bibinfo {author}
  {\bibfnamefont {G.}~\bibnamefont {Ha}}, \bibinfo {author} {\bibfnamefont
  {J.}~\bibnamefont {Power}}, \bibinfo {author} {\bibfnamefont
  {E.}~\bibnamefont {Wisniewski}},\ and\ \bibinfo {author} {\bibfnamefont
  {J.}~\bibnamefont {Rosenzweig}},\ }\bibfield  {title} {\bibinfo {title}
  {Multileaf collimator for real-time beam shaping using emittance exchange},\
  }\href@noop {} {\bibfield  {journal} {\bibinfo  {journal} {arXiv preprint
  arXiv:2107.00125}\ } (\bibinfo {year} {2021}{\natexlab{a}})}\BibitemShut
  {NoStop}%
\bibitem [{\citenamefont {Majernik}\ \emph
  {et~al.}(2021{\natexlab{b}})\citenamefont {Majernik}, \citenamefont
  {Andonian}, \citenamefont {Rosenzweig}, \citenamefont {Roussel},
  \citenamefont {Doran}, \citenamefont {Ha}, \citenamefont {Power},\ and\
  \citenamefont {Wisniewski}}]{IPACProc}%
  \BibitemOpen
  \bibfield  {author} {\bibinfo {author} {\bibfnamefont {N.}~\bibnamefont
  {Majernik}}, \bibinfo {author} {\bibfnamefont {G.}~\bibnamefont {Andonian}},
  \bibinfo {author} {\bibfnamefont {J.}~\bibnamefont {Rosenzweig}}, \bibinfo
  {author} {\bibfnamefont {R.}~\bibnamefont {Roussel}}, \bibinfo {author}
  {\bibfnamefont {S.}~\bibnamefont {Doran}}, \bibinfo {author} {\bibfnamefont
  {G.}~\bibnamefont {Ha}}, \bibinfo {author} {\bibfnamefont {J.}~\bibnamefont
  {Power}},\ and\ \bibinfo {author} {\bibfnamefont {E.}~\bibnamefont
  {Wisniewski}},\ }\bibfield  {title} {\bibinfo {title} {Arbitrary longitudinal
  pulse shaping with a multi-leaf collimator and emittance exchange},\
  }\href@noop {} {\bibfield  {journal} {\bibinfo  {journal} {Proceedings of
  12th Int. Particle Accelerator Conf. (IPAC’21), Campinas, Brazil}\ }
  (\bibinfo {year} {2021}{\natexlab{b}})}\BibitemShut {NoStop}%
\bibitem [{\citenamefont {Jordan}\ and\ \citenamefont
  {Williams}(1994)}]{jordan1994design}%
  \BibitemOpen
  \bibfield  {author} {\bibinfo {author} {\bibfnamefont {T.~J.}\ \bibnamefont
  {Jordan}}\ and\ \bibinfo {author} {\bibfnamefont {P.~C.}\ \bibnamefont
  {Williams}},\ }\bibfield  {title} {\bibinfo {title} {The design and
  performance characteristics of a multileaf collimator},\ }\href@noop {}
  {\bibfield  {journal} {\bibinfo  {journal} {Physics in Medicine \& Biology}\
  }\textbf {\bibinfo {volume} {39}},\ \bibinfo {pages} {231} (\bibinfo {year}
  {1994})}\BibitemShut {NoStop}%
\bibitem [{\citenamefont {Boyer}\ \emph {et~al.}(1992)\citenamefont {Boyer},
  \citenamefont {Ochran}, \citenamefont {Nyerick}, \citenamefont {Waldron},\
  and\ \citenamefont {Huntzinger}}]{boyer1992clinical}%
  \BibitemOpen
  \bibfield  {author} {\bibinfo {author} {\bibfnamefont {A.~L.}\ \bibnamefont
  {Boyer}}, \bibinfo {author} {\bibfnamefont {T.~G.}\ \bibnamefont {Ochran}},
  \bibinfo {author} {\bibfnamefont {C.~E.}\ \bibnamefont {Nyerick}}, \bibinfo
  {author} {\bibfnamefont {T.~J.}\ \bibnamefont {Waldron}},\ and\ \bibinfo
  {author} {\bibfnamefont {C.~J.}\ \bibnamefont {Huntzinger}},\ }\bibfield
  {title} {\bibinfo {title} {Clinical dosimetry for implementation of a
  multileaf collimator},\ }\href@noop {} {\bibfield  {journal} {\bibinfo
  {journal} {Medical physics}\ }\textbf {\bibinfo {volume} {19}},\ \bibinfo
  {pages} {1255} (\bibinfo {year} {1992})}\BibitemShut {NoStop}%
\bibitem [{\citenamefont {Ge}\ \emph {et~al.}(2014)\citenamefont {Ge},
  \citenamefont {O’Brien}, \citenamefont {Shieh}, \citenamefont {Booth},\
  and\ \citenamefont {Keall}}]{ge2014toward}%
  \BibitemOpen
  \bibfield  {author} {\bibinfo {author} {\bibfnamefont {Y.}~\bibnamefont
  {Ge}}, \bibinfo {author} {\bibfnamefont {R.~T.}\ \bibnamefont {O’Brien}},
  \bibinfo {author} {\bibfnamefont {C.-C.}\ \bibnamefont {Shieh}}, \bibinfo
  {author} {\bibfnamefont {J.~T.}\ \bibnamefont {Booth}},\ and\ \bibinfo
  {author} {\bibfnamefont {P.~J.}\ \bibnamefont {Keall}},\ }\bibfield  {title}
  {\bibinfo {title} {Toward the development of intrafraction tumor deformation
  tracking using a dynamic multi-leaf collimator},\ }\href@noop {} {\bibfield
  {journal} {\bibinfo  {journal} {Medical physics}\ }\textbf {\bibinfo {volume}
  {41}},\ \bibinfo {pages} {061703} (\bibinfo {year} {2014})}\BibitemShut
  {NoStop}%
\bibitem [{\citenamefont {Adelmann}\ \emph {et~al.}(2009)\citenamefont
  {Adelmann}, \citenamefont {Kraus}, \citenamefont {Ineichen}, \citenamefont
  {Russell}, \citenamefont {Bi},\ and\ \citenamefont
  {Yang}}]{adelmann2009object}%
  \BibitemOpen
  \bibfield  {author} {\bibinfo {author} {\bibfnamefont {A.}~\bibnamefont
  {Adelmann}}, \bibinfo {author} {\bibfnamefont {C.}~\bibnamefont {Kraus}},
  \bibinfo {author} {\bibfnamefont {Y.}~\bibnamefont {Ineichen}}, \bibinfo
  {author} {\bibfnamefont {S.}~\bibnamefont {Russell}}, \bibinfo {author}
  {\bibfnamefont {Y.}~\bibnamefont {Bi}},\ and\ \bibinfo {author}
  {\bibfnamefont {J.}~\bibnamefont {Yang}},\ }\bibfield  {title} {\bibinfo
  {title} {The object oriented parallel accelerator library (opal), design,
  implementation and application},\ }in\ \href@noop {} {\emph {\bibinfo
  {booktitle} {Proceedings of the 2009 Particle Accelerator Conference}}}\
  (\bibinfo {year} {2009})\BibitemShut {NoStop}%
\bibitem [{\citenamefont {Ha}\ \emph {et~al.}(2016{\natexlab{b}})\citenamefont
  {Ha}, \citenamefont {Cho}, \citenamefont {Gai}, \citenamefont {Kim},
  \citenamefont {Namkung},\ and\ \citenamefont
  {Power}}]{PhysRevAccelBeams.19.121301}%
  \BibitemOpen
  \bibfield  {author} {\bibinfo {author} {\bibfnamefont {G.}~\bibnamefont
  {Ha}}, \bibinfo {author} {\bibfnamefont {M.~H.}\ \bibnamefont {Cho}},
  \bibinfo {author} {\bibfnamefont {W.}~\bibnamefont {Gai}}, \bibinfo {author}
  {\bibfnamefont {K.-J.}\ \bibnamefont {Kim}}, \bibinfo {author} {\bibfnamefont
  {W.}~\bibnamefont {Namkung}},\ and\ \bibinfo {author} {\bibfnamefont {J.~G.}\
  \bibnamefont {Power}},\ }\bibfield  {title} {\bibinfo {title}
  {Perturbation-minimized triangular bunch for high-transformer ratio using a
  double dogleg emittance exchange beam line},\ }\href
  {https://doi.org/10.1103/PhysRevAccelBeams.19.121301} {\bibfield  {journal}
  {\bibinfo  {journal} {Phys. Rev. Accel. Beams}\ }\textbf {\bibinfo {volume}
  {19}},\ \bibinfo {pages} {121301} (\bibinfo {year}
  {2016}{\natexlab{b}})}\BibitemShut {NoStop}%
\bibitem [{\citenamefont {Chiadroni}\ \emph {et~al.}(2017)\citenamefont
  {Chiadroni}, \citenamefont {Alesini}, \citenamefont {Anania}, \citenamefont
  {Bacci}, \citenamefont {Bellaveglia}, \citenamefont {Biagioni}, \citenamefont
  {Bisesto}, \citenamefont {Cardelli}, \citenamefont {Castorina}, \citenamefont
  {Cianchi} \emph {et~al.}}]{Chiadroni2017}%
  \BibitemOpen
  \bibfield  {author} {\bibinfo {author} {\bibfnamefont {E.}~\bibnamefont
  {Chiadroni}}, \bibinfo {author} {\bibfnamefont {D.}~\bibnamefont {Alesini}},
  \bibinfo {author} {\bibfnamefont {M.}~\bibnamefont {Anania}}, \bibinfo
  {author} {\bibfnamefont {A.}~\bibnamefont {Bacci}}, \bibinfo {author}
  {\bibfnamefont {M.}~\bibnamefont {Bellaveglia}}, \bibinfo {author}
  {\bibfnamefont {A.}~\bibnamefont {Biagioni}}, \bibinfo {author}
  {\bibfnamefont {F.}~\bibnamefont {Bisesto}}, \bibinfo {author} {\bibfnamefont
  {F.}~\bibnamefont {Cardelli}}, \bibinfo {author} {\bibfnamefont
  {G.}~\bibnamefont {Castorina}}, \bibinfo {author} {\bibfnamefont
  {A.}~\bibnamefont {Cianchi}}, \emph {et~al.},\ }\bibfield  {title} {\bibinfo
  {title} {Beam manipulation for resonant plasma wakefield acceleration},\
  }\href@noop {} {\bibfield  {journal} {\bibinfo  {journal} {Nuclear
  instruments and methods in physics research section a: accelerators,
  spectrometers, detectors and associated equipment}\ }\textbf {\bibinfo
  {volume} {865}},\ \bibinfo {pages} {139} (\bibinfo {year}
  {2017})}\BibitemShut {NoStop}%
\bibitem [{\citenamefont {Barber}(2015)}]{BarberThesis}%
  \BibitemOpen
  \bibfield  {author} {\bibinfo {author} {\bibfnamefont {S.~K.}\ \bibnamefont
  {Barber}},\ }\href@noop {} {\emph {\bibinfo {title} {Plasma wakefield
  experiments in the quasi nonlinear regime}}}\ (\bibinfo  {publisher}
  {University of California, Los Angeles},\ \bibinfo {year} {2015})\BibitemShut
  {NoStop}%
\bibitem [{\citenamefont {Rosenzweig}\ \emph {et~al.}(2010)\citenamefont
  {Rosenzweig}, \citenamefont {Andonian}, \citenamefont {Ferrario},
  \citenamefont {Muggli}, \citenamefont {Williams}, \citenamefont {Yakimenko},\
  and\ \citenamefont {Xuan}}]{rosenzweig2010plasma}%
  \BibitemOpen
  \bibfield  {author} {\bibinfo {author} {\bibfnamefont {J.}~\bibnamefont
  {Rosenzweig}}, \bibinfo {author} {\bibfnamefont {G.}~\bibnamefont
  {Andonian}}, \bibinfo {author} {\bibfnamefont {M.}~\bibnamefont {Ferrario}},
  \bibinfo {author} {\bibfnamefont {P.}~\bibnamefont {Muggli}}, \bibinfo
  {author} {\bibfnamefont {O.}~\bibnamefont {Williams}}, \bibinfo {author}
  {\bibfnamefont {V.}~\bibnamefont {Yakimenko}},\ and\ \bibinfo {author}
  {\bibfnamefont {K.}~\bibnamefont {Xuan}},\ }\bibfield  {title} {\bibinfo
  {title} {Plasma wakefields in the quasi-nonlinear regime},\ }in\ \href@noop
  {} {\emph {\bibinfo {booktitle} {AIP Conference Proceedings}}},\ Vol.\
  \bibinfo {volume} {1299}\ (\bibinfo {organization} {American Institute of
  Physics},\ \bibinfo {year} {2010})\ pp.\ \bibinfo {pages}
  {500--504}\BibitemShut {NoStop}%
\bibitem [{\citenamefont {Andonian}\ \emph {et~al.}(2011)\citenamefont
  {Andonian}, \citenamefont {Williams}, \citenamefont {Wei}, \citenamefont
  {Niknejadi}, \citenamefont {Hemsing}, \citenamefont {Rosenzweig},
  \citenamefont {Muggli}, \citenamefont {Babzien}, \citenamefont {Fedurin},
  \citenamefont {Kusche} \emph {et~al.}}]{andonian2011resonant}%
  \BibitemOpen
  \bibfield  {author} {\bibinfo {author} {\bibfnamefont {G.}~\bibnamefont
  {Andonian}}, \bibinfo {author} {\bibfnamefont {O.}~\bibnamefont {Williams}},
  \bibinfo {author} {\bibfnamefont {X.}~\bibnamefont {Wei}}, \bibinfo {author}
  {\bibfnamefont {P.}~\bibnamefont {Niknejadi}}, \bibinfo {author}
  {\bibfnamefont {E.}~\bibnamefont {Hemsing}}, \bibinfo {author} {\bibfnamefont
  {J.}~\bibnamefont {Rosenzweig}}, \bibinfo {author} {\bibfnamefont
  {P.}~\bibnamefont {Muggli}}, \bibinfo {author} {\bibfnamefont
  {M.}~\bibnamefont {Babzien}}, \bibinfo {author} {\bibfnamefont
  {M.}~\bibnamefont {Fedurin}}, \bibinfo {author} {\bibfnamefont
  {K.}~\bibnamefont {Kusche}}, \emph {et~al.},\ }\bibfield  {title} {\bibinfo
  {title} {Resonant excitation of coherent cerenkov radiation in dielectric
  lined waveguides},\ }\href@noop {} {\bibfield  {journal} {\bibinfo  {journal}
  {Applied Physics Letters}\ }\textbf {\bibinfo {volume} {98}},\ \bibinfo
  {pages} {202901} (\bibinfo {year} {2011})}\BibitemShut {NoStop}%
\bibitem [{\citenamefont {Yakimenko}\ \emph {et~al.}(2019)\citenamefont
  {Yakimenko}, \citenamefont {Alsberg}, \citenamefont {Bong}, \citenamefont
  {Bouchard}, \citenamefont {Clarke}, \citenamefont {Emma}, \citenamefont
  {Green}, \citenamefont {Hast}, \citenamefont {Hogan}, \citenamefont {Seabury}
  \emph {et~al.}}]{yakimenko2019facet}%
  \BibitemOpen
  \bibfield  {author} {\bibinfo {author} {\bibfnamefont {V.}~\bibnamefont
  {Yakimenko}}, \bibinfo {author} {\bibfnamefont {L.}~\bibnamefont {Alsberg}},
  \bibinfo {author} {\bibfnamefont {E.}~\bibnamefont {Bong}}, \bibinfo {author}
  {\bibfnamefont {G.}~\bibnamefont {Bouchard}}, \bibinfo {author}
  {\bibfnamefont {C.}~\bibnamefont {Clarke}}, \bibinfo {author} {\bibfnamefont
  {C.}~\bibnamefont {Emma}}, \bibinfo {author} {\bibfnamefont {S.}~\bibnamefont
  {Green}}, \bibinfo {author} {\bibfnamefont {C.}~\bibnamefont {Hast}},
  \bibinfo {author} {\bibfnamefont {M.}~\bibnamefont {Hogan}}, \bibinfo
  {author} {\bibfnamefont {J.}~\bibnamefont {Seabury}}, \emph {et~al.},\
  }\bibfield  {title} {\bibinfo {title} {Facet-ii facility for advanced
  accelerator experimental tests},\ }\href@noop {} {\bibfield  {journal}
  {\bibinfo  {journal} {Physical Review Accelerators and Beams}\ }\textbf
  {\bibinfo {volume} {22}},\ \bibinfo {pages} {101301} (\bibinfo {year}
  {2019})}\BibitemShut {NoStop}%
\bibitem [{\citenamefont {Scheinker}\ \emph {et~al.}(2021)\citenamefont
  {Scheinker}, \citenamefont {Huang},\ and\ \citenamefont
  {Taylor}}]{scheinker2021extremum}%
  \BibitemOpen
  \bibfield  {author} {\bibinfo {author} {\bibfnamefont {A.}~\bibnamefont
  {Scheinker}}, \bibinfo {author} {\bibfnamefont {E.-C.}\ \bibnamefont
  {Huang}},\ and\ \bibinfo {author} {\bibfnamefont {C.}~\bibnamefont
  {Taylor}},\ }\bibfield  {title} {\bibinfo {title} {Extremum seeking-based
  control system for particle accelerator beam loss minimization},\ }\href@noop
  {} {\bibfield  {journal} {\bibinfo  {journal} {IEEE Transactions on Control
  Systems Technology}\ } (\bibinfo {year} {2021})}\BibitemShut {NoStop}%
\bibitem [{\citenamefont {Duris}\ \emph {et~al.}(2020)\citenamefont {Duris},
  \citenamefont {Kennedy}, \citenamefont {Hanuka}, \citenamefont {Shtalenkova},
  \citenamefont {Edelen}, \citenamefont {Baxevanis}, \citenamefont {Egger},
  \citenamefont {Cope}, \citenamefont {McIntire}, \citenamefont {Ermon} \emph
  {et~al.}}]{duris2020bayesian}%
  \BibitemOpen
  \bibfield  {author} {\bibinfo {author} {\bibfnamefont {J.}~\bibnamefont
  {Duris}}, \bibinfo {author} {\bibfnamefont {D.}~\bibnamefont {Kennedy}},
  \bibinfo {author} {\bibfnamefont {A.}~\bibnamefont {Hanuka}}, \bibinfo
  {author} {\bibfnamefont {J.}~\bibnamefont {Shtalenkova}}, \bibinfo {author}
  {\bibfnamefont {A.}~\bibnamefont {Edelen}}, \bibinfo {author} {\bibfnamefont
  {P.}~\bibnamefont {Baxevanis}}, \bibinfo {author} {\bibfnamefont
  {A.}~\bibnamefont {Egger}}, \bibinfo {author} {\bibfnamefont
  {T.}~\bibnamefont {Cope}}, \bibinfo {author} {\bibfnamefont {M.}~\bibnamefont
  {McIntire}}, \bibinfo {author} {\bibfnamefont {S.}~\bibnamefont {Ermon}},
  \emph {et~al.},\ }\bibfield  {title} {\bibinfo {title} {Bayesian optimization
  of a free-electron laser},\ }\href@noop {} {\bibfield  {journal} {\bibinfo
  {journal} {Physical review letters}\ }\textbf {\bibinfo {volume} {124}},\
  \bibinfo {pages} {124801} (\bibinfo {year} {2020})}\BibitemShut {NoStop}%
\end{thebibliography}%

\end{document}